# Extraordinarily Large Permittivity Modulation in Zinc Oxide for Dynamic Nanophotonics

Soham Saha[1,2], Aveek Dutta[1,2], Clayton DeVault[6], Benjamin T. Diroll[3], Richard D. Schaller[3,4], Zhaxylyk Kudyshev[1,2], Xiaohui Xu[1,2], Alexander Kildishev[1,2], Vladimir M. Shalaev[1,2], and Alexandra Boltasseva[1,2] *

[1]*School of Electrical and Computer Engineering, Purdue University, West Lafayette, IN 47907, USA*

[2]*Birck Nanotechnology Center, Purdue University, West Lafayette, IN 47907, USA*

[3]*Argonne National Laboratory, Illinois*

[5]*Department of Chemistry, Northwestern University, Evanston, Illinois*

[6]*Harvard University, MA, USA*

*\*aeb@purdue.edu*

**Abstract:** The dielectric permittivity of a material encapsulates the essential physics of light-matter interaction into the material's local response to optical excitation. Dynamic, photo-induced modulation of the permittivity can enable an unprecedented level of control over the phase, amplitude, and polarization of light. Therefore, the detailed dynamic characterization of technology-relevant materials with substantially tunable optical properties and fast response times is a crucial step in the realization of tunable optical devices. This work reports on the extraordinarily large permittivity changes in zinc oxide thin films (up to -3.6 relative change in the real part of the dielectric permittivity at 1600 nm wavelength) induced by optically generated free carriers. We demonstrate broadband reflectance modulation up to 70% in metal-backed oxide mirrors at the telecommunication wavelengths, with picosecond-scale relaxation times. The epsilon near zero points of the films can be dynamically shifted from 8.5 µm to 1.6 µm by controlling the pump fluence. Finally, we show that the modulation can be selectively enhanced at specific wavelengths employing metal-backed ZnO disks while maintaining picosecond-scale switching times. This work provides insights into the free-carrier assisted permittivity modulation in zinc oxide and could enable the realization of novel dynamic devices for beam-steering, polarizers, and spatial light modulators.

1. **Introduction**

The dielectric permittivity of a material determines the local response of the material to optical excitation. Thus, controlling the permittivity of a material enables control over the amplitude, phase, and polarization of light interacting with it. Recently, an emerging class of tailorable and dynamically tunable nanophotonic materials, namely, transparent conducting oxides (TCOs), has enabled unprecedent control over the optical response via processing and post-processing means such as eletrical and optical excitation[1–4]. Spectral control over the material permittivity by doping has led to the development of epsilon-near-zero[5] (ENZ) enhanced optical phenomena such as extraordinarily large reflectance modulation and nonlinearity enhancements[6,7]. In addition to the realization of highly tunable optical response in thin films, spatial and periodic control of the permittivity in multilayer structures is key to the design of novel metasurfaces or metamaterials for wavefront shaping[8,9]. Recent advances in static metamaterials have demonstrated an unprecedented level of manipulation on the properties of light, with demonstrations of polarization control[10,11], sensing[12], super-resolution imaging[13,14], nonlinear optics[15], and quantum optics[16].

Dynamic tuning of the material permittivity could drive several optical technologies to the next step by enabling control over the optical properties of such nanostructures in real-time and providing an additional degree of freedom in light manipulation. Active control over the phase, amplitude, and polarization of light could enable technologies such as ultrafast signal transfer and beam-steering, and at the same time, open new directions in fundamental optical research in the fields of optical time-reversal[17], nonreciprocal optics[18], and photonic time-crystals[19].

Transparent conducting oxides have been widely utilized in tunable waveguides and metasurfaces because of their tailorable optical properties[4], established deposition methods[20], and large laser damage thresholds[21]. Optically induced modulation in TCOs has the advantage of achieving fast, significant changes in the refractive index spanning hundreds of nanometers, depending on the skin-depth of the pump light. The switching speed is determined by the relaxation time of the free-carriers, which can range from hundreds of picoseconds to sub-picoseconds[22–24]. Ultrafast modulation of reflectance has been demonstrated employing the inter-band excitation of electrons in Fabry-Perot cavities[25], epsilon-near-zero substrates[22,26], and Berreman-metasurfaces[27]. Extraordinarily large reflectance modulation has been demonstrated both in the visible and the mid-IR ranges via interband or intraband pumping[23,28]. However, to date, a systematic investigation of the optical free-carrier induced permittivity changes of these oxides with different pump-fluences remains mostly unexplored.

In this paper, we perform a systematic characterization of free-carrier assisted permittivity modulation in zinc oxide films and nanostructures. The paper is divided into five sections. In Section 1, we demonstrate broadband all-optical reflectance modulation of up to 70% in metal-backed dielectric mirrors, with relaxation times of ~20 ps at the telecommunication wavelength of 1300 nm. In Section 2, we show that the modulation results from large permittivity changes in the unpatterned ZnO film. The real part of the relative permittivity of the films decreases with the increasing pump fluence, reaching saturation around 23 mJ/cm$^2$, while the imaginary part increases with increasing pump-fluence. The permittivity modulation reported here is on par with ENZ-enhanced nonlinearities observed in indium doped tin oxide[3] or aluminum-doped zinc oxide[29]. In Section 3, we determine the physical origins of the permittivity modulation. In Section 4, we demonstrate that the temporal response of the mirrors is determined by the faster response time in zinc oxide. Finally, in Section 5, we further demonstrate TiN-backed ZnO nanostructures that selectively enhance the reflection modulation at specific wavelengths and polarizations. A reflection modulation of 55% is shown for s-polarized light at 1100 nm with a pump-fluence of 7.6 mJ/cm$^2$, with picosecond relaxation times. The large permittivity modulation of undoped zinc oxide establishes it as a viable candidate for the design of ultrafast, dynamically tunable optical devices for amplitude, phase, and polarization modulation, and calls for the exploration of the dynamic properties of other low-loss undoped oxides for applications in dynamic nanophotonics. The techniques employed here can be extended to the study of other doped and undoped conducting oxides, enabling a better understanding of the carrier dynamics and the permittivity modulation in TCOs for active nanophotonic applications.

## 2. Ultrafast, broadband reflection modulation with ~20 ps relaxation times

The ideal properties of all-optical modulators include fast switching response, deep modulation, and large bandwidth[30]. We chose low-loss, undoped zinc oxide (ZnO) for our studies. ZnO has been widely explored in microelectronic and nanophotonic applications ranging from photodiodes[31] to single-photon sources[32], with established fabrication procedures, and detailed experimental studies of its steady-state optical properties with various dopants[33], making it the ideal material for the study.

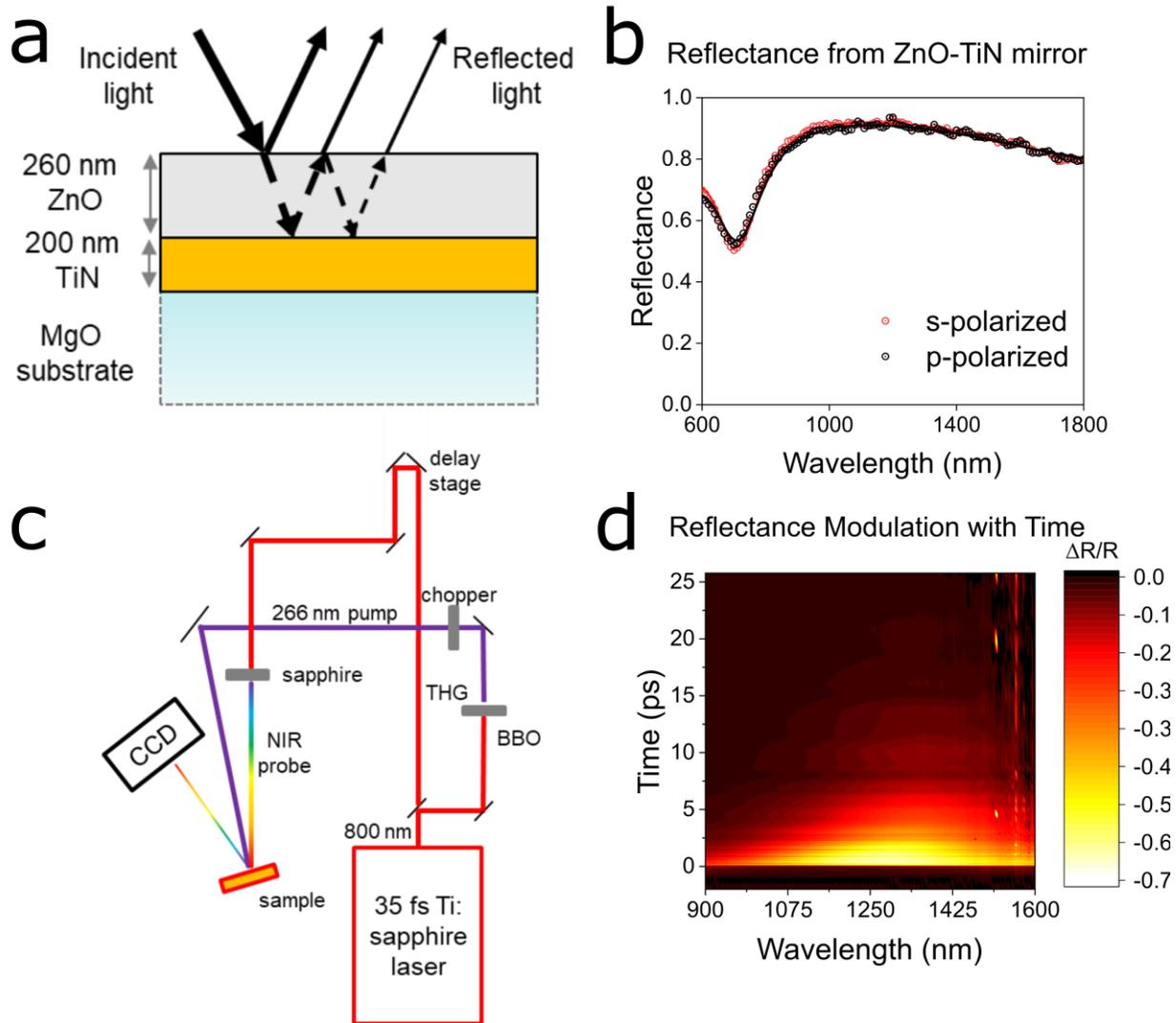

*Figure 1 (a) Schematic of the TiN-ZnO mirror. The incident probe light makes multiple passes in the dielectric ZnO as it is reflected by the TiN-ZnO and ZnO-air interfaces. (b) Steady-state reflectance of s and p polarized light measured at 20º angle of incidence. The dots represent the measured reflectance and the solid lines represent the simulated reflectance computed using the optical properties of the layers via the transfer matrix method. (c) Pump-probe setup. (d) Color map showing the temporal (vertical-axis) and spectral (horizontal axis) dynamics of the reflection modulation under an interband pump at a pump fluence of 31.6 mJ/cm$^2$. The maximum modulation is around -70%, the negative sign corresponding to a decrease in the reflectance in the pumped state.*

We fabricated a metal-backed dielectric mirror to demonstrate all-optical switching in planar nanocavities with deep, broadband modulation and picosecond response times. Figure 1(a) shows the schematic of the mirror. An optically thick, 200 nm thick film of epitaxial TiN was grown on MgO, followed by the deposition of polycrystalline ZnO with a thickness of 260 nm. The high thermal and laser damage threshold[34] of titanium nitride allowed us to study the modulation of the devices up to the saturation fluences without risking the structural and morphological integrity of the films. Supporting Information 1 and 2 have a detailed overview of the fabrication process and the optical properties, respectively.

Reflectance measurements at a 20° angle of incidence show the structure to have Fabry-Perot like resonances in the optical and NIR frequency regime for both *s*- and *p*-polarized light (Fig. 1(b)). We performed variable angle spectroscopic ellipsometry (VASE) at 50° and 70° angles of incidence to extract the permittivities of the TiN and ZnO films. The data from 600 nm – 1800 nm was fitted using a Drude-Lorentz model, comprising one Drude term and one Lorentz term. The relative permittivity, represented as a complex number, is given by the following equation,

$$\varepsilon = \varepsilon_1 + i\varepsilon_2 = \varepsilon_\infty + \frac{A_1}{E_1^2 - (\hbar\omega)^2 - iB_1\hbar\omega} - \frac{A_0}{(\hbar\omega)^2 + iB_0\hbar\omega} \quad (1)$$

where $\varepsilon_1$ and $\varepsilon_2$ are the real and imaginary parts of the dielectric function, $\hbar$ is the reduced Plank constant, $\omega$ is the probe angular frequency, and $\varepsilon_\infty$ is an additional offset; $A_0$ is the square of the plasma frequency in electron volts, $B_0$ is the damping factor of the Drude oscillator; $A_1$, $B_1$, and $E_1$ are the amplitude, broadening, and the center energy of the Lorentz oscillator respectively. Table S2.1 in Supporting Information 2 presents the Drude-Lorentz model parameters used to fit the ellipsometry data. In the near-infrared (NIR) regime studied (900-1600 nm), ZnO acts as a low-loss dielectric with a positive relative $\varepsilon_1$, and low losses ($\varepsilon_2$), as shown in Fig. S1(a). TiN acts like a metal, with a negative $\varepsilon_1$ and higher losses ($\varepsilon_2$) (Fig. S1(d)).

To study the reflectance-modulation via interband pumping, we performed pump-probe experiments with a 266 nm pump and a broadband probe from 900-1400 nm, with pump-fluences ranging from 1.3 mJ/cm² to 31.6 mJ/cm²/pulse. Figure 1(c) shows the experimental setup. The reflectance modulation, $\Delta R/R$, is defined as,

$$\frac{\Delta R}{R} = \left(\frac{R_{initial} - R_{final}}{R_{initial}}\right) \quad (2)$$

Figure 1(d) illustrates the color map of the reflectance modulation observed at a pump-fluence of 31.6 mJ/cm². A broadband modulation of over 70% is observed at the maximum temporal overlap of the pump and the probe, centered around the telecommunication wavelength of 1300 nm. The modulation has a fast decay time, falling close to zero under 25 ps.

Figure 2(a) illustrates the reflectance modulation mechanism. An interband pump excites electrons from the valence band of zinc oxide to the conduction band and changes the permittivity of the films by Drude-dispersion. The increase in free carrier density changes the plasma frequency, decreasing the refractive index of the ZnO. The increased number of free carriers also increases scattering due to collisions of the free-electrons with imperfections in the lattice, resulting in a subsequent increase in the damping and increases the absorption in the dielectric layer. As a result, the reflection of the layer decreases, and absorbance increases. The free-carriers eventually recombine through direct or defect-assisted recombination, returning the reflectance to its original unexcited state.

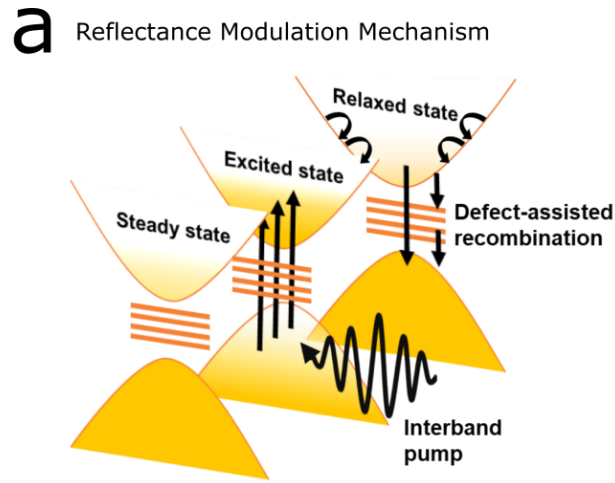

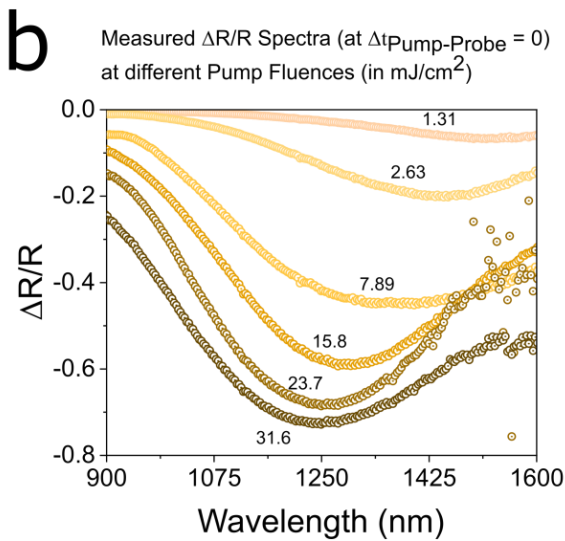

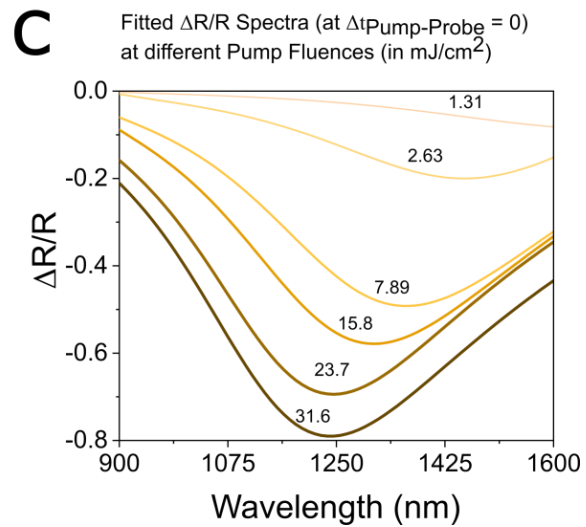

*Figure 2 (a) Mechanism of interband transition induced permittivity modulation; an interband pump excites electrons from the conduction to the valence band, changing the permittivity through Drude-dispersion; as the electrons recombine, assisted by defects in the lattice, the material returns to its relaxed state. (b) Experimentally obtained reflectance spectra at different pump powers at the peak modulation. The peak is seen to blue shift with increasing pump-fluence. (c) Simulated ΔR/R obtained by fitting the plasma frequency and the damping parameters of the ZnO film.*

The magnitude of the reflectance modulation increases with the increasing pump-fluence (Fig. 2c). A peak value of -70% is attained at a pump fluence of 31.6 mJ/cm², where the negative sign corresponds to a decrease in the reflection upon pumping.

The reflectance modulation at the maximum pump-probe temporal overlap can be modeled by changing the Drude dispersion term in the Drude-Lorentz oscillator model via an increase of the value of $A_0$ and $B_0$ in Equation 1. The experimentally obtained reflectances (Fig. 2(c)) match well with the fitted graphs thus obtained(Fig. 2(d)). The extracted Drude parameters are presented in Table 1. The reflection modulation is also polarization-independent at a low angle of incidence, and probing with p-polarized light shows similar values of reflection modulation and spectrum at similar pump-fluences (Supporting Information Fig. S3(a,b)).

The maximum modulation magnitude achieved here is similar in magnitude to the extraordinarily large modulation seen in aluminum-doped zinc oxide samples operating near epsilon-near-zero, which saturates at 75% ($\Delta T/T$) at a pump fluence of 15mJ/cm$^2$ [26]. Even without any ENZ enhancement, optical-pumping in undoped ZnO demonstrates broadband, high-amplitude, deep modulation utilizing simple, lithography-free metal-dielectric cavities. This effect makes it possible to design free-space all-optical switches with the material without the need of doping adjectment, ENZ point tailoring, and lithography.

3. **Magnitude and limits of permittivity and refractive index changes in ZnO with interband doping**

The permittivity values of the photo-excited ZnO were extracted from the plasma frequencies and the Drude damping factor at each fluence. The real part of the permittivity becomes more negative with increasing pump fluence and saturates around 23 mJ/cm$^2$ (Fig. 3(a)). The imaginary part corresponding to the losses increases with increasing fluence throughout the range of fluences studied. Undoped ZnO has an ENZ point at 8.5 µm. After pumping, the optically-induced ENZ values can be blue-shifted down to 1600 nm. The imaginary part of the permittivity (losses) of these optically-induced ENZ values ranges from 0.5-1.0 (Fig. 3(b)), which is comparable to that seen in metal-doped conducting oxides such as AZO, GZO, and ITO[3,35].

As the permittivity of ZnO decreases with increasing pump power, the reflectance spectra of the mirror blue-shift. There is also more absorption in the films, causing a dip in the overall reflection. The relative change in the modulation depends on both the final and the initial reflectance at a particular wavelength, gradually blue-shifting with increasing pump-fluence, as shown in Fig. 3(c).

Prior work by Alam et al. in ITO demonstrated extraordinarily large refractive index change of ITO working near the epsilon near zero points (1.2 at 200 GW/cm$^2$ peak power, at 1200 nm wavelength)[3]. In our work, a $\Delta n$ value of -1 is achieved at 1650 nm, at a peak power of 223 GW/cm$^2$ without the ENZ enhancement or metal doping (Fig. 3(d)). The optically induced modulation persists across the entire depth of the ZnO film, as opposed to the voltage induced modulation with similar changes, that occur across a few nanometers of the accumulation layer[36].

It should be noted here that the sign of both the permittivity change and the losses is the opposite of what is observed with intraband pumping, as reported by Caspani et al. in their work on AZO with intraband pumping[37]. The increase in losses due to Drude dispersion is another crucial factor that needs to be considered when modeling the optical properties of dielectric nanostructures under a pump. The findings establish that extraordinary changes in the permittivity can be achieved in ZnO by pumping it to its saturation, well below its laser damage threshold. The large permittivity changes could enable the design of novel, all-optical amplitude modulators, phase-shifters, and polarization rotators. The optically induced ENZ properties at NIR wavelengths also may enable the study of interesting nonlinear optical phenomena[7] with dynamically tunable ENZ values without dependence on metal dopants.

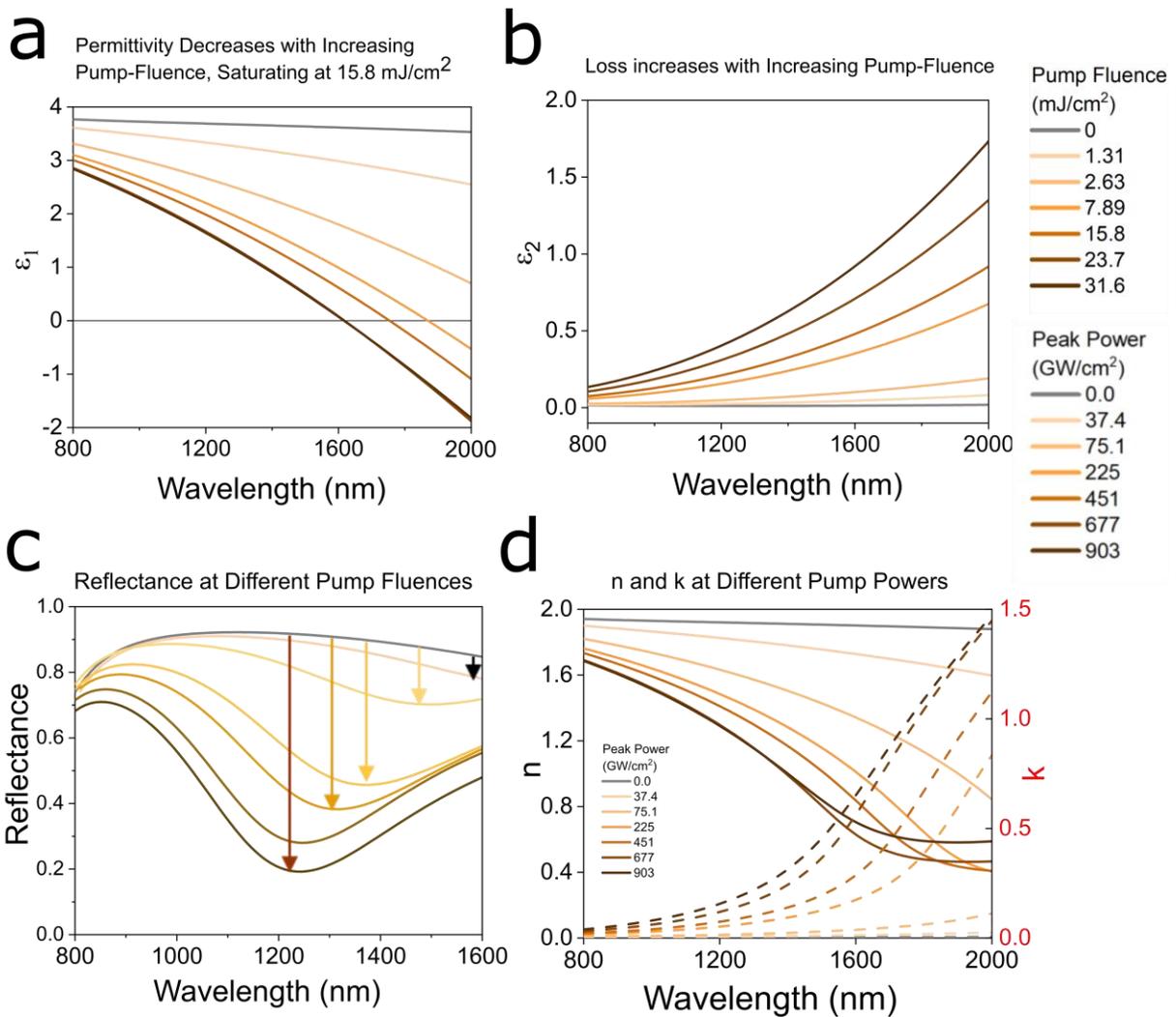

*Figure 3. (a) Real and imaginary part of the permittivities of the ZnO film. (b) (c) Reflectance values of the mirrors with different pumps. (d) Real and imaginary part of the ZnO film refractive index under different pump fluence.*

## 4. Mechanism of permittivity changes in ZnO

This section investigates how the permittivity change in ZnO is affected by the photoexcited carrier concentration, their effective masses, and the Drude damping factor.

The equation (1) describing the relative permittivity of ZnO in the Drude Lorentz model can be rewritten by separating the real and the imaginary parts.

The real part of the permittivity, $\varepsilon_1$, which largely determines the refractive index, is given by

$$\varepsilon_1 = \varepsilon_b - \frac{A_0}{(\hbar\omega)^2 + iB_0\hbar\omega} \quad (3)$$

Where $\varepsilon_b$ comprises the net contribution of offset $\varepsilon_\infty$ and the Lorentz oscillator, which plays a more substantial role in the optical properties of the ZnO from the ultraviolet to visible range. The plasma frequency squared term, $A_0$, depends on the free carrier density, $N$, and the effective mass of the electrons and holes ($m_e$ and $m_h$),

$$A_0 = \frac{Ne^2}{\varepsilon_0 m^* m_0}\hbar^2 \quad (4)$$

where $m^*$ is the total effective mass of the generated electrons and the holes given by $m^* = (m_e^{-1} + m_h^{-1})^{-1}$, $m_0$ is the electronic mass, and $\hbar$ is the reduced Planck constant.

The imaginary part, $\varepsilon_2$, determining the loss, is given by

$$\varepsilon_2 = \frac{A_0 B_0}{(\hbar\omega)((\hbar\omega)^2 + B_0^2)} \quad (5)$$

From Equations (3)-(5), it follows that as carrier density increases, the plasma frequency increases, the permittivity decreases, and the loss increases.

As light with energy above the bandgap impinges on the surface, several mechanisms take place in parallel. The increased number of photoexcited carriers increase the density of both electrons and holes ($N$ increases from a steady-state value of ~$10^{19}$ cm$^{-3}$ to a maximum value close to $6 \times 10^{20}$ cm$^{-3}$). The plasma frequency term ($A_0$) increases from 0.08 eV$^2$ to 2 eV$^2$ at a pump fluence of 15.8 mJ/cm$^2$ and saturates near 2.2 eV with higher pump fluencies (Fig. S4 (a)). The saturation is a result of the lower branches of the conduction band getting filled and the bandgap widening, resulting in fewer carriers moving from the valence to the conduction band (Burstein-Moss effect)[38]. There is also a depletion of the donor states that takes place with an increase in the number of photoexcited carriers. This effect results in the saturation of the free-carrier density with increasing pump fluence.

As the photoexcited carriers populate higher regions away from the band edges, the average effective masses of both electrons and holes increases, due to the non-parabolic nature of the band of the ZnO[39,40]. This slows down the increase in plasma frequency with the fluence, causing the real part of the permittivity to reach saturation. The increased number of carriers and the lattice heating with the optical pumping also results in an increased number of collisions between the photoexcited carriers and the grain boundaries, and the Drude damping factor $B_0$ increases to up to 2.5 times its initial value. The effective mass of the photoexcited carriers can also be estimated from eq. (4) (details in Supporting Information 4). Table 1 shows the plasma frequency, the effective mass, the estimated photoexcited carrier concentration, and the Drude damping.

*Table 1. Drude parameters, carrier concentrations and effective mass vs pump-fluence.*

| Pump-fluence (mJ/cm$^2$) | $A_0$ (eV$^2$) | $B_0$ (eV) | N ($\times 10^{20}$cm$^{-3}$) | $m^*$ |
|---|---|---|---|---|
| 0 | 0.085 | 0.040 | 0.140 | 0.17[41] |
| 1.31 | 0.437 | 0.040 | 0.559 | 0.17 |
| 2.63 | 1.11 | 0.038 | 1.11 | 0.18 |
| 7.89 | 1.59 | 0.097 | 3.34 | 0.14 |
| 15.8 | 2.02 | 0.19 | 6.67 | 0.29 |
| 23.7 | 2.15 | 0.15 | 6.67 | 0.48 |
| 31.6 | 2.22 | 0.19 | 6.67 | 0.43 |

The Drude damping factor and the effective mass both increase with increasing pump fluence. These trends match closely with the trends seen in ZnO films heavily doped with increasing aluminum concentrations. In Al-doped ZnO, as the concentration of carriers increased from 0.92x10$^{20}$ cm$^{-3}$ to 4.01x10$^{20}$ cm$^{-3}$, the mobility showed a decrease of 2.6 times[40]. In our experiments, as the carrier concentration increases from 1.4 x10$^{19}$ to 6x10$^{20}$ cm$^{-3}$, the Drude damping factor increases by 2.5 times. Similar trends are also seen in the increase in the effective mass of the carriers, which increase with an increase in carrier concentration in both optically pumped (this work) and aluminum-doped ZnO[40].

The net result of all the effects is that the real part of the permittivity reaches a minimum at a pump-fluence of 23.7 mJ/cm$^2$, while subsequent increase in the pump-fluence only results in an increase in the imaginary part (losses).

5. **Relaxation of photoexcited carriers in individual ZnO and TiN films, and the metal-dielectric mirror**

After photoexcitation, carriers eventually recombine via radiative and non-radiative channels, returning the permittivity of the material to its steady-state values. Thus, the minimum attainable switching time in interband modulation largely depends on the relaxation time of the carriers. The relaxation is best approximated with a single exponential model, given by

$$\left(\frac{\Delta R(t)}{R}\right)(t) = A\, e^{\frac{-t}{\tau}} + B \quad (6)$$

where $A$ is the amplitude of the modulation, $t$ is the pump-probe delay time, $\tau$ is the decay constant, and $B$ is an offset attributed to thermal energy dissipation due to lattice heating.

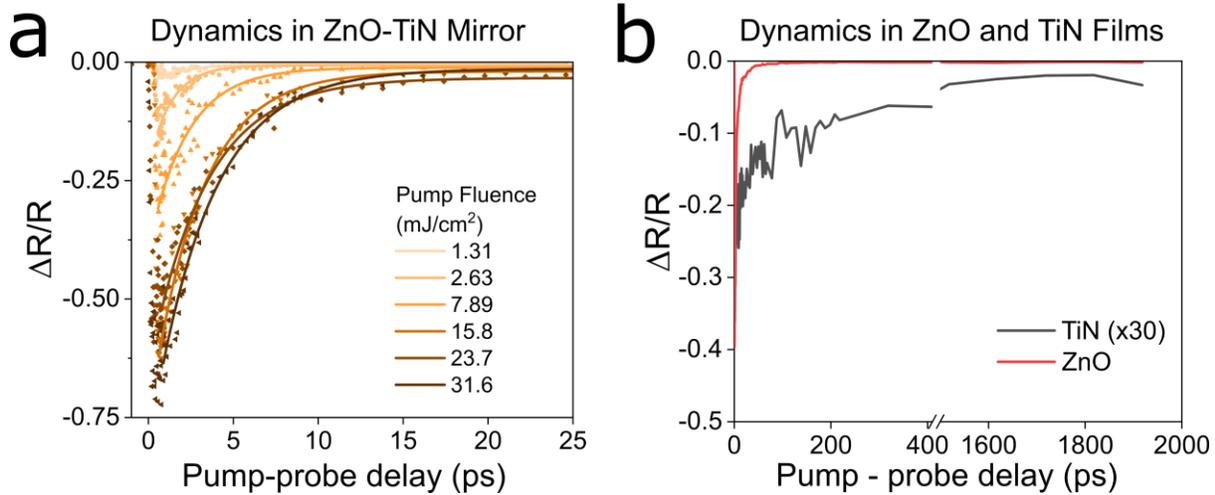

*Figure 4 (a) Decay constants for the ZnO-TiN mirrors for different pump-fluences at 1300 nm; (b) Normalized reflectance modulation for a 2000 ps scan for TiN film on MgO, and ZnO film on fused silica at a pump fluence of 30 mJ/cm$^2$, showing the order of magnitude difference in the relaxation times.*

The relaxation constant for the change is ZnO is around 3 ps (Table S4.2) and is fairly invariant with an increase in the pump fluence (Fig. 4(a)). The constant increases slightly with increasing pump fluence (corresponding to an increase in the photoexcited carrier concentration). This increase points to the dominant relaxation mechanism to be defect-assisted Shockley-Read-Hall mechanism[22,42], aided by defects in the grain boundaries and oxygen-related defects present in the zinc oxide. The relaxation time in undoped zinc oxide is approximately an order of magnitude slower than that in aluminum-doped zinc oxide, which follows a similar relaxation process aided by the greater density of defect states in the doped oxides[22]. The relaxation time in cadmium oxide also has been reported to show a significant decrease with increasing dopant-induced defect density[42]. The fast response is followed by a slower tail of 50-100 ps that corresponds to the gradual cooling of the lattice[23].

*Effect of the slow relaxation dynamics of the TiN on the response time of the ZnO-TiN mirror:* It is important to note here that TiN has been showed to have a very slow optical response time ranging in the order of nanoseconds[43,44]. Figure 4(b) displays the normalized response of a ZnO film on fused silica versus that of a bare TiN film on MgO at a pump fluence of 30 mJ/cm$^2$. The ZnO film reflectance modulation has a magnitude of 0.4 and falls to zero under 100 ps. TiN, on the other hand, has a much lower modulation (~0.01), enhanced 30x, and a slower response time lasting up to a few nanoseconds[43]. The larger reflectance change in ZnO compared to TiN under the same pump-fluence causes the ZnO film to "dictate" the response of the ZnO-TiN mirror.

Furthermore, the finite element method (FEM) simulation with COMSOL employing the optical properties of TiN and ZnO at 266 nm shows that 93% of the incident light is absorbed in the zinc oxide cladding and around 1.5% in the TiN layer. Any change in the TiN caused by the pump reaching the TiN layer results in a negligible change in the TiN property. As a result, the dynamics of the metal-dielectric mirror is largely determined by the dynamics of the ZnO film.

It is also noteworthy that the static or the dynamic responses of the measured spot does not show any sign of change even at the highest laser fluence used, underpinning the robustness of the ZnO and TiN building blocks. The experiment shows that broadband, large modulation can be achieved from metal-dielectric systems operating even with undoped TCOs without the ENZ-enhancement.

**6. Enhancing the modulation depth at specific wavelengths with multimode, polarization selective resonators**

In this section, we utilize a resonant structure employing the same materials to show modulation enhancement at targeted wavelengths. The unit cell for our metasurface comprises a zinc oxide disk on a metal reflector (inset of Fig. 5(a)). Such metal-backed disks yield efficient scattering in the far-field and offer highly directional emission at visible and infrared wavelengths in the near field[45]. The resonant wavelength of such disks can be tailored by changing the disk diameter or height. The large field-enhancements in similar metal-backed dielectric metasurfaces have been utilized for lasing[46]. For our project, they offer the benefit of supporting multiple resonant modes which are simultaneously tuned.

ZnO nanodisks with 600 nm diameter, 200 nm height, with a 1000 nm period support two resonant modes at 900 nm and 1270 nm (Fig. 5(a)-black curve) at a 20º angle of incidence for s-polarized light. The structure is polarization sensitive, as seen from the different reflectance spectra for *p*-polarized light under the same configurations (Fig. 5(a) - red curve). Figure 5(b) shows the mode-profiles of the hybrid-plasmonic modes for both *s*- and *p*-polarized lights at the reflectance minima.

Through FEM simulations, we investigated the effect of optical pumping on the reflectance modulation of *s*-polarized light. Under excitation, the permittivity of the ZnO changes, and the resonances blue-shift. Using the optical properties derived from the previous section for different pump fluencies, the reflectivity of the structure at the maximum temporal overlap of pump-probe can be computed. The refractive index of the ZnO decreases with increasing pump-fluence, resulting in a blue-shift of the resonant wavelength with increasing pump-fluence (Fig. 5(c)). The maximum modulation occurs at the steady-state (unpumped) resonant wavelength, where the steady-state reflectance value is minimum. At the same wavelength, the reflectance increases with increasing pump-fluence (shown by the vertical arrow). The change eventually saturates at a pump-fluence of 7.89 mJ/cm$^2$, where the reflectance of the device reaches its maximum value. Further increase in pump power blue-shifts the resonance dip, but the reflectance at 1270 nm saturates, and modulation reaches a maximum value.

Figure 5(d) shows the reflectance modulation with different pump fluences. It is interesting to note here that the reflectance modulation magnitude is controlled mostly by the minimum reflectance at the resonant wavelength. The closer it is to perfect absorption, the greater is the reflectance modulation. An ideal dielectric disk on metal can demonstrate ΔR/R values of up to 1600%. Unlike the ZnO-TiN mirrors discussed in the earlier sections, in this case, the wavelength of maximum modulation is pinned at the steady-state resonance minimum.

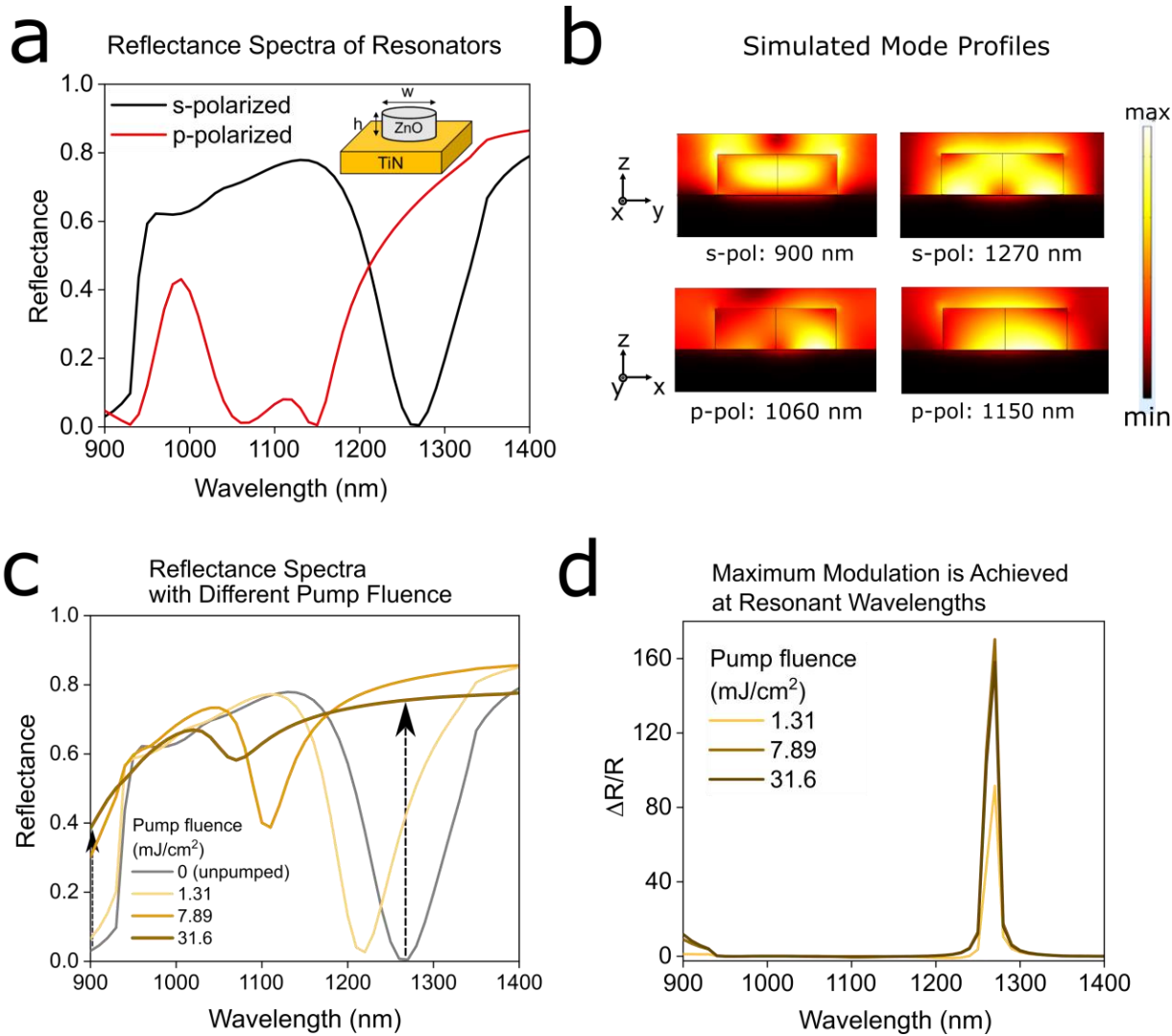

*Figure 5 (a) Reflectance of s and p-polarized light from the disks (at 20° angle of incidence) from the resonators. Inset shows the schematic of the dielectric disk on the metal backreflector. (b) Mode profiles showing the normalized electric fields in a cross-section parallel to the electric field. (c) The simulated reflectance of s-polarized light from the metasurface with different pump-fluences. The arrows indicate points of maximum reflection modulation (d) Peak reflection modulation vs wavelength at different pump fluence.*

Although a significant amount of light reaches the TiN layer in this resonator, FEM simulations show that any optically-induced change in the titanium nitride underlayer results in negligible changes in the reflectance modulation of the hybrid plasmonic nanodisks (Supporting Information 5).

*Experimental observation of reflectance modulation:*

For our investigations, we aimed fabricated 200 nm disks with 600-nanometer width, with a periodicity of 1000 nm. The resonators were fabricated via standard electron beam lithography and liftoff. Due to imperfections in the lift-off process, the disks had a slight taper (Fig. 6(a)), and the average width was around 580 nm, measured by Atomic Force Microscopy (Supporting Information 7).

The measured reflectance spectrum for s-polarized light shows two dips in the reflection at 930 and 1100 nm at a 20º angle of incidence. Figure 6(b) shows the simulated and measured reflectance spectra of the tapered nanodisks, showing a good match. The simulated mode-profile is similar to that of the ideal disks (Fig. 6(c)). Transient pump-probe spectroscopy shows large reflection modulations at 930 and 1100 nm, corresponding to these resonant dips (Fig. 6(d)).

Reflection modulations of 10% are apparent even at low pump fluences of 0.95 mJ/cm$^2$, with a maximum modulation of 55% observed at 1100 nm for a pump fluence of 7.6 mJ/cm$^2$. The modulation increases with increasing pump-fluence and does not show spectral shifts with increasing pump power. Similar results are observed for p-polarized light, with a maximum reflectance modulation of 20% (Fig. S6(a,b)).

As shown by the SEM and AFM images, the topology of the disks is quite rough, and the height and width of the nanodisks vary across the substrate due to fabrication imperfections. The roughness and the width variations cause the resonant dips of individual nanodisks to vary, resulting in an overall decrease in the modulation from the simulated value. Incorporating the height variations and the roughness into the simulations result in a good match between the reflectance modulation for both s- and p-polarized light (Supporting Information 7).

The temporal dynamics of the reflection modulation can be fit using a single time-constant model (Table S4.2), similar to the previous section. The corresponding decay time here is around 3 ps for all pump fluences (Table S7.1), similar to that for the unpatterned ZnO on TiN.

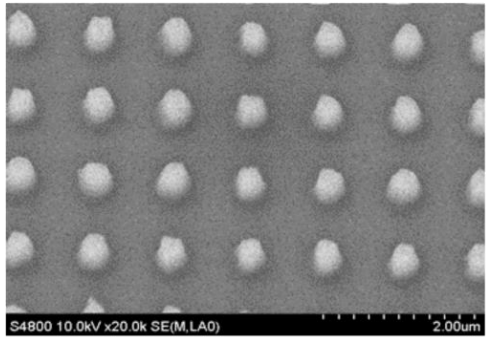
a SEM Image of ZnO Nanodisk

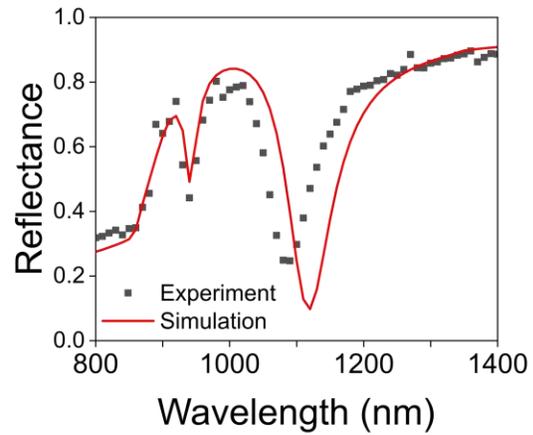
b Simulated and Measured Reflectance Spectrum of ZnO Nanodisks on TiN for s-polarized light

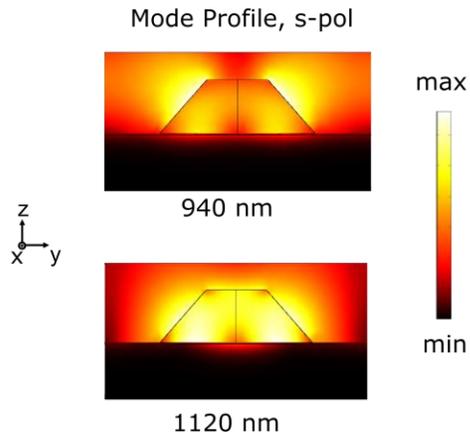
c Mode Profile, s-pol
940 nm
1120 nm

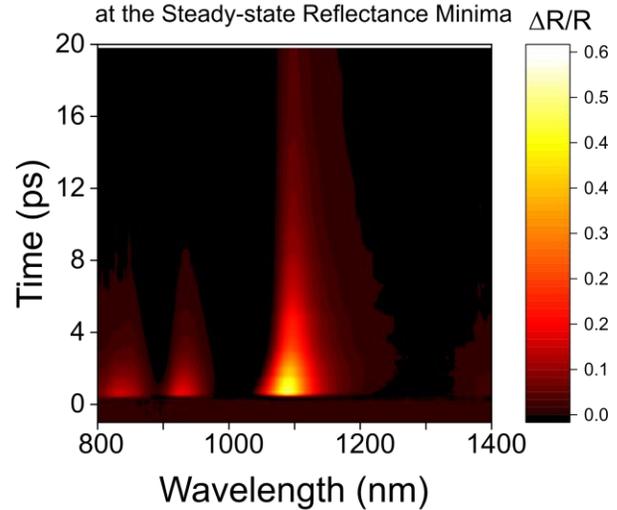
d Reflectance Modulation is Enhanced at the Steady-state Reflectance Minima

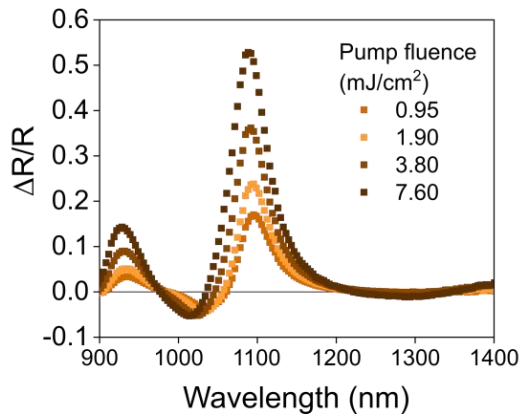
e Experimental ΔR/R for s-polarized light from the ZnO nanodisks on TiN

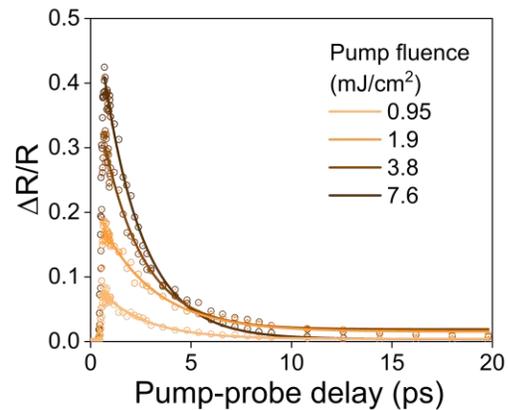
f Transient ΔR/R at 1080 nm from the ZnO Nanodisks on TiN

*Figure 6 (a) SEM of the fabricated disks with the tapered profile (b) Experimental reflectance of s-polarized light at an angle of 20o measured by spectroscopic ellipsometry (black dots). The red line shows the simulated result for a 450nm wide, 200 nm high disk. (c) The tapered disks have similar mode profiles to the vertical ones near resonance. (d) Color map showing the reflection modulation dynamics: the modulation reaches its peak at the steady-state reflectance minima, corresponding to the resonance wavelengths of 1100 and 940 nm (e) The modulation increases with pump-fluence, reaching a peak of 55% at 7.6 mJ/cm$^2$, and does not shift with pump fluence (f) The resonators show similar relaxation rates of around 3 ps.*

The experiments conclude that the optically-induced reflectance modulation can be spectrally engineered by the design of metal-backed ZnO nanodisk resonators. The modulation is comparable to prior work demonstrated with ITO nanopillars[23], ENZ metasurfaces[26], and Mie-resonant metasurfaces[25]. It is possible to tailor the structures to attain perfect absorption and further enhance the modulation depth at similar pump fluences.

## 7. Conclusion

In summary, this work establishes undoped ZnO as a robust, tunable, high-power tolerant material for ultrafast tunable optics. The broadband, large-amplitude (70% at 23.7 mJ/cm$^2$ pump fluence), and fast (~20 ps) reflection modulation may be utilized in lithography-free, all-optical switches operating at the telecommunication wavelengths. The unity-order changes in the refractive index of ZnO can be utilized in the design of all-optical metasurfaces for phase modulation and beam steering. The reflectance modulation can be selectively enhanced by simple structures such as the disk-on-metal resonator designed, without compromising the device speed. An ultrafast, optically tunable metasurface is demonstrated to have a high reflectance modulation of 55% at 1100 nm at a pump fluence of 7.6 mJ/cm$^2$. Optimization of fabrication procedures for ZnO nanostructures promises to further decrease the power requirements for optical switches and enhance the modulation with maximum theoretical values of 1600%. The techniques employed in this work can be extended to the study of other doped and undoped conducting oxides for ultrafast nanophotonic applications. Overall, this work benefits material research involving the fast modulation of the amplitude, phase, and polarization of light for dynamic photonic applications.

## 8. Data availability

The data that support all plots within this paper is available from the corresponding author upon reasonable request.

## 9. Author contributions

S. Saha conceived the idea, designed and supervised the project. A. Dutta grew the ZnO films. C. Devault, S. Saha, and B. Diroll performed the measurements. X. Xu performed AFM characterization. Z. Kudyshev and A. Kildishev aided with simulations. R. Schaller contributed to data acquisition and discussion. R. Schaller, A. Kildishev, V. Shalaev, A. Boltasseva contributed to paper writing and funding acquisition. All authors contributed to commenting on the manuscript.

## 10. Acknowledgements

This work was performed in part at the Center for Nanoscale Materials, a U.S. Department of Energy Office of Science User Facility, and supported by the U.S. Department of Energy, Office of Science, under Contract No. DE-AC02-06CH11357. Part of this work was also supported by the Air Force Office of Scientific Research through Award Nos. FA9550-18-1-0002, and FA9550-19-S-0003. The authors acknowledge the support of the U.S. Department of Energy, Office of Basic Energy Sciences, Division of Materials Sciences and Engineering, under Award No. DE-SC0017717 (sample preparation), Office of Naval Research Grant N00014-18-1-2481, and DARPA/DSO Extreme Optics and Imaging (EXTREME) program HR00111720032.

**Supporting Information (SI)**

*SI 1. Methods*

**TiN and ZnO film growth**

The titanium nitride films were deposited on magnesium oxide substrates using DC magnetron sputtering at 800°C. A 99.995% pure titanium target of 2 in. the diameter was used. The DC power was set at 200 W. To maintain a high purity of the grown films, the chamber was pumped down to $3\times10^{-8}$ Torr before deposition and backfilled to 5 mTorr during the sputtering process with argon. The throw length was kept at 20 cm, ensuring a uniform thickness of the grown TiN layer throughout the 1 cm by 1cm MgO substrate. After heating, the pressure increased to $1.2\times10^{-7}$ Torr. An argon-nitrogen mixture at a rate of 4 sccm/6 sccm was flowed into the chamber. The deposition rate was 2.2 Å $\text{min}^{-1}$.

ZnO was deposited on the TiN films by pulsed laser deposition (PLD) with a KrF excimer laser at a pump fluence of 1.5 J/cm$^2$. The chamber was pumped down to $6\times10^{-6}$ T and then backfilled to 3 mTorr with oxygen. The substrate temperature was 45°C, and the growth rate was 6 nm/min.

**ZnO-TiN resonator fabrication**

200 nm TiN was grown on MgO using the method in the previous section.

The ZnO disks are fabricated by Electron Beam Lithography (EBL), followed by ZnO deposition and lift-off. ZEP 520 A resist was spin-coated at 1000 rpm for 1 minute (with 2s ramp up and ramp down times), and post-baked at 180°C for 2 mins. EBL was done at a dose of 320 µC to form holes, followed by a development in xylene for 75s. ZnO was grown on the resist with PLD, with the recipe described in the previous section. The lift-off was carried out by immersion in ZDMAC (dimethylacetamide) for 10 min and sonication for 1 min, followed by rinsing with acetone and IPA.

**Pump-probe experiments**

The 266 nm pump is formed by the third harmonic of an 800 nm, 35 fs pulse-width Ti-sapphire laser passed through a BBO crystal. A delayed NIR supercontinuum probe was generated by focusing the 800 nm laser on a sapphire plate. The pump beam was oriented with a normal angle of incidence to the substrate, and the probe beam was oriented with an angle of incidence of ~20°.

## SI 2. Optical models of ZnO and TiN

**Spectroscopic ellipsometry and fits:** The TiN and the ZnO films were characterized using a J. A. Woollam VASE (Variable Angle Spectroscopic Ellipsometer) at angles of 50 and 70°.

From the visible to the NIR wavelengths, both films can be fitted using a Drude-Lorentz model, where the free carriers are modeled with a Drude oscillator and the net effect of bound electrons with a Lorentz term.

The relative permittivity represented as a complex number, $\varepsilon = \varepsilon_1 + i\varepsilon_2$, is given by the following equation,

$$\varepsilon = \varepsilon_1 + i\varepsilon_2 = \varepsilon_\infty - \frac{A_0}{(\hbar\omega)^2 + iB_0\hbar\omega} + \frac{A_1}{E_1^2 - (\hbar\omega)^2 - iB_1\hbar\omega}$$

where $\varepsilon_1$ and $\varepsilon_2$ are the real and imaginary parts of the dielectric function, $h$ is the Plank constant, $\hbar\omega$ is the probe energy, and $\varepsilon_\infty$ is an additional offset; $A_0$ is the square of the plasma frequency in electron volts, $B_0$ is the damping factor of the Drude oscillator; $A_1$, $B_1$, and $E_1$ are the amplitude, broadening, and the center energy of the Lorentz oscillator respectively.

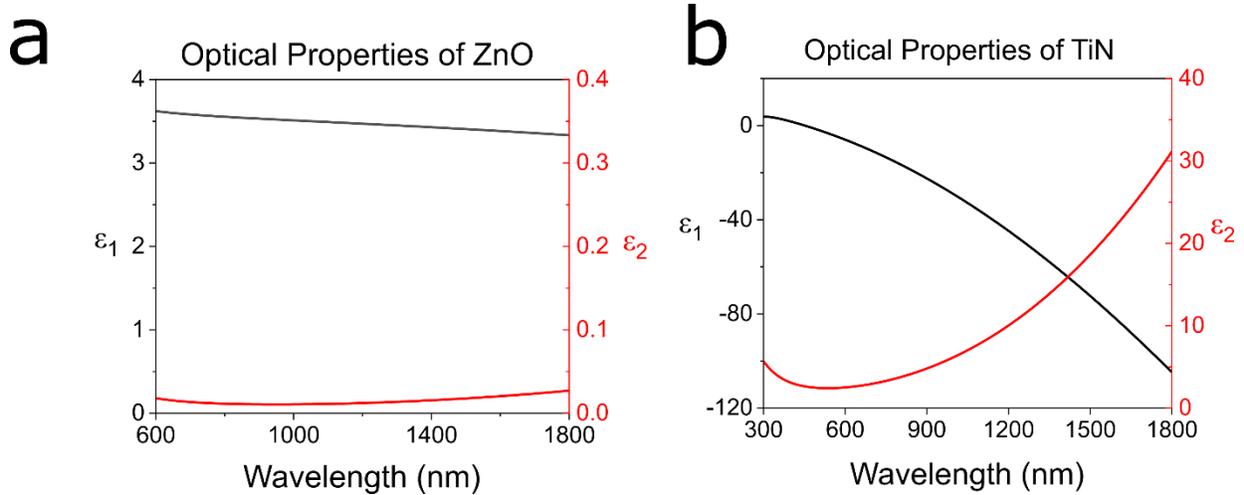

*Figure S1. (a) Optical properties comprising the real (black) and imaginary (red) relative permittivities of the as-deposited ZnO film; (b) Optical properties of the as-deposited TiN.*

Supporting Table S2.1 has the values of the Drude-Lorentz parameters, respectively, used to fit the ellipsometry data.

*Table S2.1. Drude-Lorentz Model Parameters for TiN and ZnO*

| Drude Lorentz model parameters | TiN | ZnO |
|---|---|---|
| $\varepsilon_\infty$ | 3.44 | 3.32 |
| $A_0$ (eV)$^2$ | 57.6 | 0.085 |
| $B_0$ (eV) | 0.197 | 0.0396 |
| $A_1$ (eV)$^2$ | 119.1 | 7.93 |
| $B_1$ (eV) | 3.096 | 0.317 |
| $E_1$ (eV) | 5.26 | 4.34 |
| MSE | 3.64 | 7.77 |

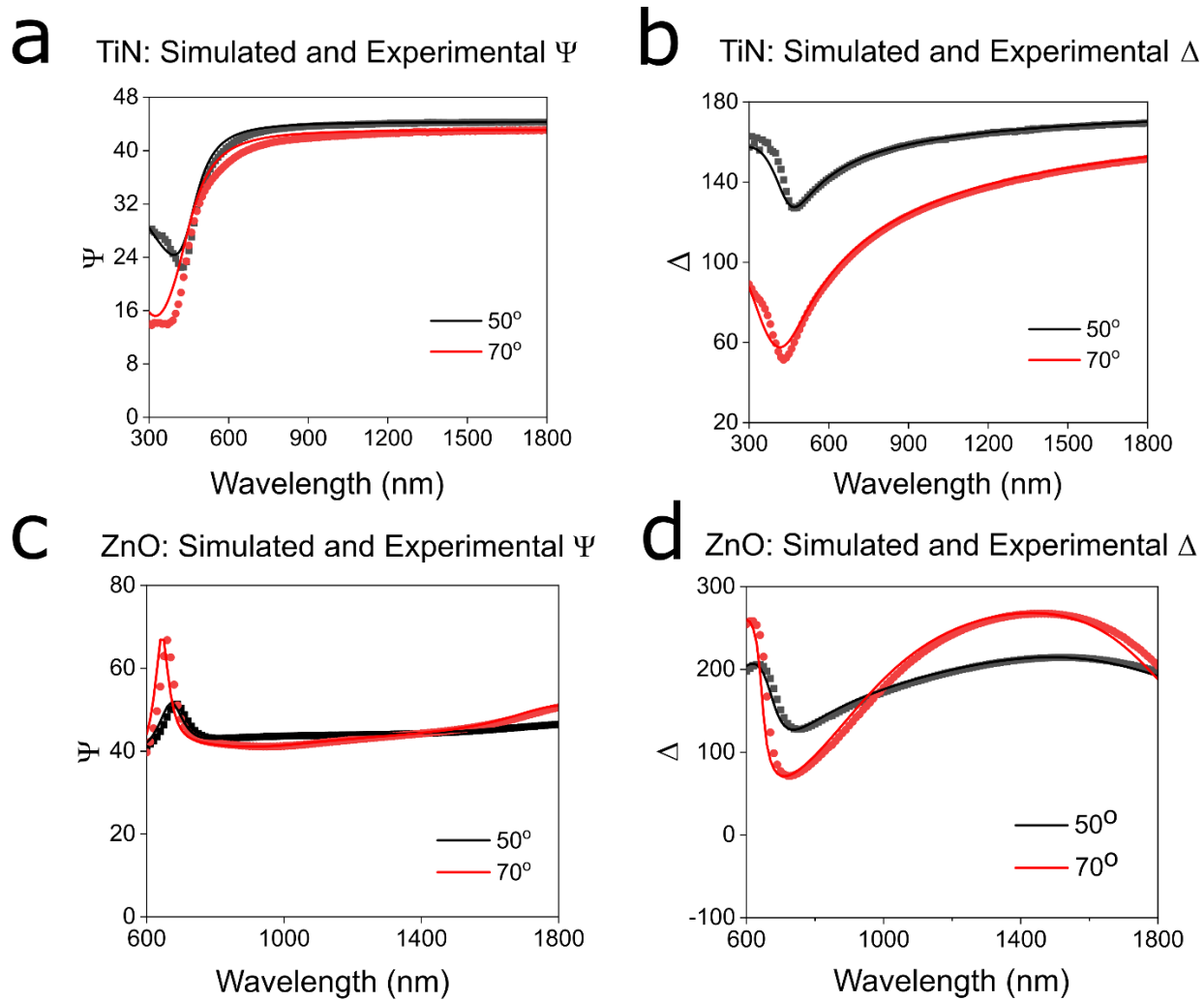

*Figure S2. Simulated and Experimental psi delta values of the TiN (a and b) and ZnO (c and d) layers from the Drude-Lorentz Model.*

### SI 3. P polarized simulation and experiments for TiN-ZnO mirror

We performed pump-probe spectroscopy at 4 pump-fluences with p-polarized light with identical configurations to that used for s-polarized light for the planar dielectric-metal mirror. The modulation shows similar traits to that of the s-polarized lights. The Drude-parameters from SI1 were used to generate the reflectance modulation spectra, showing a good match with the experimental data for both amplitude and spectral response (Fig. S3(b)).

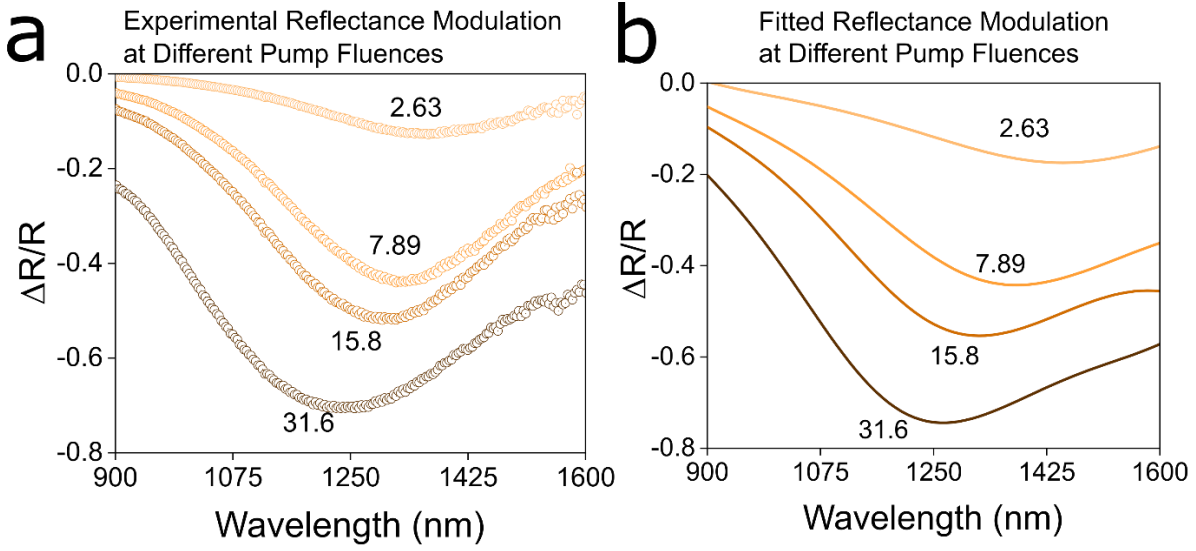

*Figure S3. (a) Reflectance modulation vs pump fluence. (b) Calculated reflectance modulation vs pump-fluence from Drude-Lorentz parameters*

### SI 4. Changes in the carrier concentration, effective mass, and damping

The number of free electrons and holes, *N* can be estimated from the formula

$$N = FA \frac{1 - e^{-\alpha t_l}}{t_l E_{\text{pump}}}$$

where $F$ is the pump fluence, $A$ is the absorbance of the layer, $t_l$ is the layer thickness, $E_{\text{pump}}$ is the energy of the 266-nm photons, and $\alpha$ is the absorbance of the zinc oxide at 266 nm.

To estimate the carrier concentration in ZnO, the near-bandgap optical properties of ZnO from ellipsometric fits is needed. We add a Tauc-Lorentz oscillator to the Drude Lorentz model to extract the optical properties of the films measured at an angle of 75°. The Drude parameters are unchanged, but the Lorentz parameters are fitted.

In the Tauc-Lorentz model[1], the imaginary part of the complex dielectric function is given by:

$$\varepsilon_2 = \frac{AE_0\,\Gamma\,(E-E_g)^2}{(E^2-E_0^2)^2+\Gamma^2 E^2} \frac{1}{E}, E > E_g,$$

$$\varepsilon_2 = 0, E \leqslant E_g$$

where the parameter $E_g$ denotes the bandgap; $E_0$ denotes the peak transition energy; $\Gamma$ is the broadening parameter, and $A$ represents the optical transition matrix elements. The real part of the dielectric function is obtained by the Kramers-Kronig integration of $\varepsilon_2$:

$$\varepsilon_1 = \varepsilon_\infty + \frac{2}{\pi} P \int_{E_g}^{\infty} \frac{\xi \varepsilon_2(\xi)}{\xi^2 - E^2} d\xi$$

where $\varepsilon_\infty$ represents the contribution of the optical transitions at higher energies and appears as an additional fitting parameter. Supporting Table S4.1 has the values of the Tauc-Lorentz parameters used to fit the ellipsometry data.

*Table S4.1. Tauc Lorentz parameters for the optical constants for the films from 250 to 500 nm.*

| TL Parameter | | Drude-Lorentz Parameters | |
|---|---|---|---|
| $\varepsilon_\infty$ | 1.98 | $A_0$ (eV)$^2$ | 0.085 |
| A (eV) | 544 | $B_0$ (eV) | 0.040 |
| $E_0$ (eV) | 3.12 | $A_1$ (eV)$^2$ | 3.600 |
| $\Gamma$ (eV) | 0.174 | $B_1$ (eV) | 0.126 |
| $E_g$ (eV) | 3.14 | $E_1$ (eV) | 3.212 |
| MSE | 20 | | |

$\alpha = 2k_0\varepsilon_2$, where $k_0 = 2\pi/\lambda$ is the free-space wavevector at 266 nm.

Table S4.1 contains the details of the above-bandgap optical property estimation of ZnO.

*Table S4.1 Drude parameters, carrier concentrations and effective mass vs Pump-fluence*

| Pump-fluence (mJ/cm$^2$) | $A_0$ (eV$^2$) | $B_0$ (eV) | N (×10$^{20}$cm$^{-3}$) | m* |
|---|---|---|---|---|
| 0 | 0.085 | 0.040 | 0.140 | 0.17[2] |
| 1.31 | 0.437 | 0.040 | 0.559 | 0.17 |
| 2.63 | 1.11 | 0.038 | 1.11 | 0.18 |
| 7.89 | 1.59 | 0.097 | 3.34 | 0.14 |
| 15.8 | 2.02 | 0.19 | 6.67 | 0.29 |
| 23.7 | 2.15 | 0.15 | 6.67 | 0.48 |
| 31.6 | 2.22 | 0.19 | 6.67 | 0.43 |

The changes in the reflection are fitted by changing the values of $A_0$ and $B_0$ corresponding to changes in the plasma frequency and the Drude-damping factor, respectively. The Lorentz terms are left unperturbed, as the changes in the Lorentz oscillators have a stronger effect in the lower wavelength range in the visible, as opposed to the longer wavelengths, where the optical properties are more affected by the free-carriers.

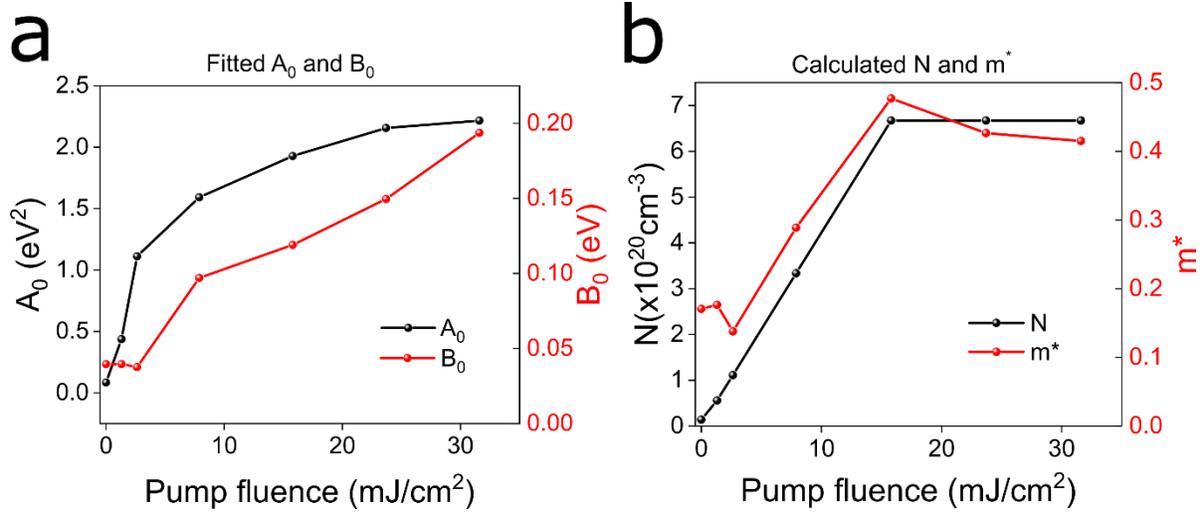

*Figure S4 (a) Plasma frequency (black curve) obtained from fitting the reflectance spectra, seen to saturate near 2 eV with increasing pump fluence; the Drude damping coefficient (red curve) increases with increasing pump fluence (b) Calculated photoexcited electron and hole concentrations (black curve). Above 15.8 mJ/cm², we estimated the carrier concentration to remain unchanged with increasing pump fluence; the effective mass of the electrons and holes increases with increasing pump fluence, saturating near 0.4;*

From the estimated carrier density, and the value of the $A_0$ extracted from fitting the reflection modulation graphs, we can calculate the effective mass m*

$$m^* = \left(\frac{1}{m_e} + \frac{1}{m_p}\right)^{-1} = \frac{\hbar^2 N e^2}{\varepsilon_0 A_0^2 m}$$

Figure S4b shows the effective mass of the electron-holes extracted from the fitted plasma frequencies. It should be noted that for 15.8mJ/cm² and above, we kept the N value constant, accounting for the saturable absorption of the pump. In reality, there should be a marginal increase in the carrier concentration with increasing pump even after 15.8 mJ/cm².

Few prior works exist studying the effective mass of photoexcited carriers, but the effect of doping on the free electrons' effective mass in zinc oxide has been thoroughly studied. In Al-doped ZnO with large carrier concentrations, as the concentration of carriers increased from $0.92 \times 10^{20}$ cm$^{-3}$ to $4.01 \times 10^{20}$ cm$^{-3}$, the mobility showed a decrease of 2.6 times[3]. In our experiments, as the carrier concentration increases from $1.4 \times 10^{19}$ to $6.7 \times 10^{20}$ cm$^{-3}$, the Drude damping factor increases by 2.5 times. This is similar in magnitude to the Damping factor increase with increased carrier concentrations in heavily doped Al:ZnO investigated by Romanyuk et al[3].

The extracted effective mass from our experiments also shows a very similar trend of increasing with free-carrier concentration as that observed in heavily doped AZO. The mass increases with increasing pump fluences reaching a peak value of around 0.4. For high pump-fluences where both light and heavy holes are excited, assuming $m_h \gg m_e$ in

ZnO, the $m_{eff}$ will approximately equal the effective mass of free electrons in ZnO. From our estimates, the effective mass saturates at around 0.4, which is similar to that of Al:ZnO at similar electron concentrations[3], giving further credibility to the fits.

The increased number of photoexcited carriers also results in an increased Drude damping in the films (Fig S2.2 c). This can be understood as the increased number of carriers contribute to an increase in the electron-electron, electron-phonon, and electron-lattice scattering.

It should be noted here that because of Fermi smearing induced by hot carriers, there may also be an associated shift in the Lorentz parameters because of the change in the amplitude and the broadening of the Lorentz parameters, but the effects of that would be more prominent in the visible wavelength region. However, changing the Lorentz parameters in our model for the ZnO result in only marginal improvements in the fits.

*Table S4.2. Time constants for the hybrid plasmonic resonator at different pump fluences*

| Pump fluence (mJ/cm$^2$) | Time-constant (ps) |
| --- | --- |
| 1.31 | 2.65 |
| 2.63 | 1.32 |
| 7.89 | 2.25 |
| 15.8 | 2.92 |
| 23.7 | 3.34 |

***SI 5. Effect of changing TiN the back-reflector properties in the hybrid plasmonic resonator***

Transient reflection measurement on bare TiN on the same sample shows a weak reflection modulation of around 1% at a pump fluence of 30 mJ/cm$^2$ (Fig. S5(a)). From the transient response of the TiN reflection modulation at the maximum pump-probe overlap, we extracted the plasma frequency and damping factors of the TiN under the pump. The plasma frequency showed a negligible change, while the damping factor increased slightly. This is expected, as the number of the photoexcited carrier generated is several orders of magnitude smaller than the intrinsic carrier concentration. This change results in an overall decrease in the reflectance across the NIR spectral range (Fig. S5(a)). Fig. S5(b) shows the simulated reflectance of the ideal nanodisk structures, considering the changes in the TiN optical properties under an interband pump. In the pumped case, when both TiN and ZnO are excited (brown dashed curve), the resonance shift is almost identical to the case where the only ZnO is excited (orange curve). Similarly, once the ZnO has reached the ground state (and TiN is still excited), the graphs for both modulated (black dashed curve) and unmodulated (grey dashed curve) overlap, showing that even though the TiN has not fallen back to the ground state, the overall reflectance modulation falls back to zero with the active ZnO, the faster of the two components.

This shows that fast active modulators can be realized in metal-dielectric nanostructures, even when one active component is significantly slower than another, provided that the faster of the two components undergo a deeper modulation.

Another thing of importance is the dependence of the switching time on the carrier relaxation rates inside the different components of the optical switch. Using two materials with switching times differing by two orders of magnitude (ZnO in with a relaxation time of 3 ps, and TiN with a relaxation time of 1 ns), we demonstrate that the switching time of free-carrier assisted modulation can be dictated by the faster of the two materials. This opens the possibility of utilizing the fast active medium for the modulation upon slow, but robust metallic components for plasmonic enhancements in nanophotonic switch design.

However, it should be noted that in practical, fast-acting switches with multiple pulses reaching the active device with picosecond scale periods, the small remnant modulation can eventually build-up and interfere with the overall device performance, making it crucial to investigate ways for active thermal management in ultrafast systems.

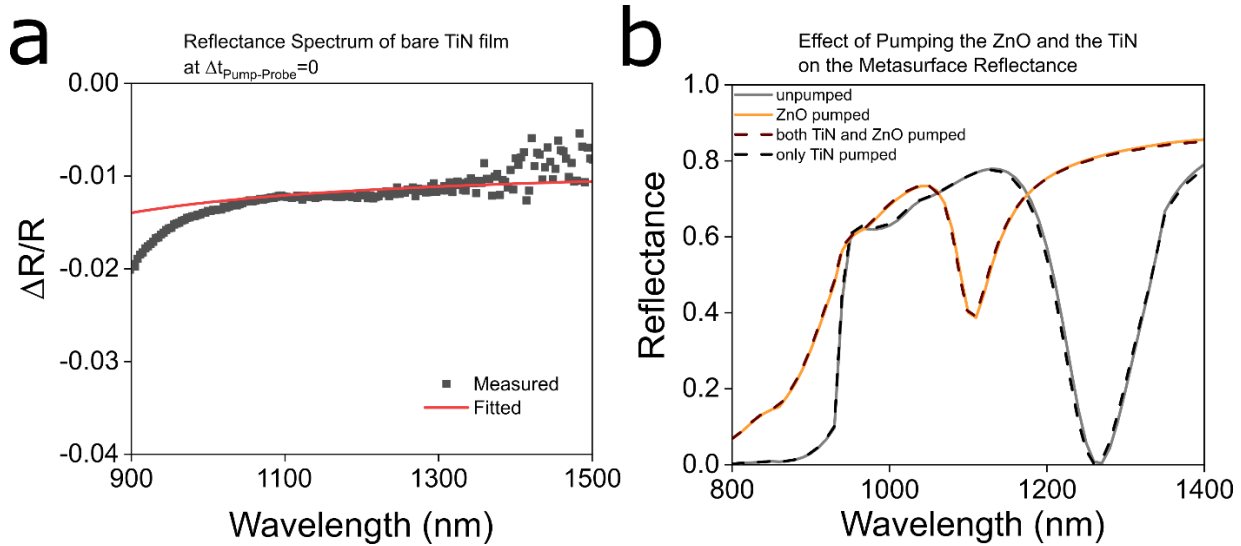

*Figure S5. (a) Spectral response at the maximum pump-probe overlap of TiN (black) and fitted curve (red) (b) Reflectance spectra of the ZnO disk on TiN structure showing the unpumped state (grey), pumped state with changes both in the ZnO and the TiN (brown dashed), pumped state with the ZnO only (orange) and the relaxed state where ZnO has returned to its ground state, but TiN is still excited (black dashed).*

*SI 6. Reflectance modulation for p-polarized light in the resonators*

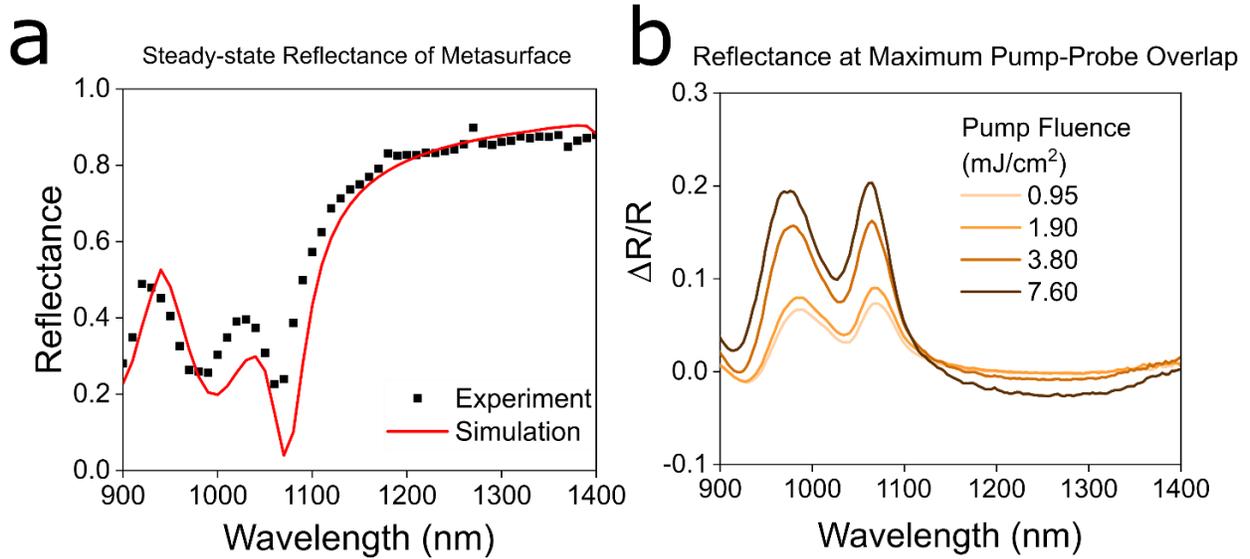

Fig S6 (a) Experimental and simulated reflectance of hybrid plasmonic resonators for p-polarized light at an angle of 20º (b) p-polarized spectrum at maximum modulation

*SI 7. Simulating reflection modulation in metal-dielectric resonators*

Atomic force microscopy on the ZnO nanodisks shows the disks to have rough sidewalls, and their heights ranged from 200 to 250 nanometers (Fig. S6(a)). Fig. S7(b,c) is the experimentally obtained reflectance spectra at the maximum pump-probe temporal overlap for *s*- and *p*-polarized light respectively. To capture the effect of the roughness and this height variation, the AFM topology was imported to the simulation tool Lumerical FDTD Solutions, and the reflectance of individual nanodisks in the unpumped and the pumped states was simulated from the parameters in Table S4.1 at the pump fluence of 7.89 mJ/cm$^2$. Because of the imperfections, the reflectance of individual nanodisks can vary, resulting in a variance in the reflectance modulation between individual nanoantennae. Fig. S7(c,d) show the experimentally observed modulation at a pump fluence of 7.6 mJ/cm$^2$ for both polarizations, while Fig. S7(d) shows the average reflectance modulation of four arbitrarily chosen nanodisks with varying roughness profiles with heights near 200 nm, in the spot illuminated by the pump. The simulated results show a good qualitative match with the experimental values, with a slight spectral shift.

Table 7.1 shows the experimentally extracted time constants at 1100 nm of the reflectance modulations.

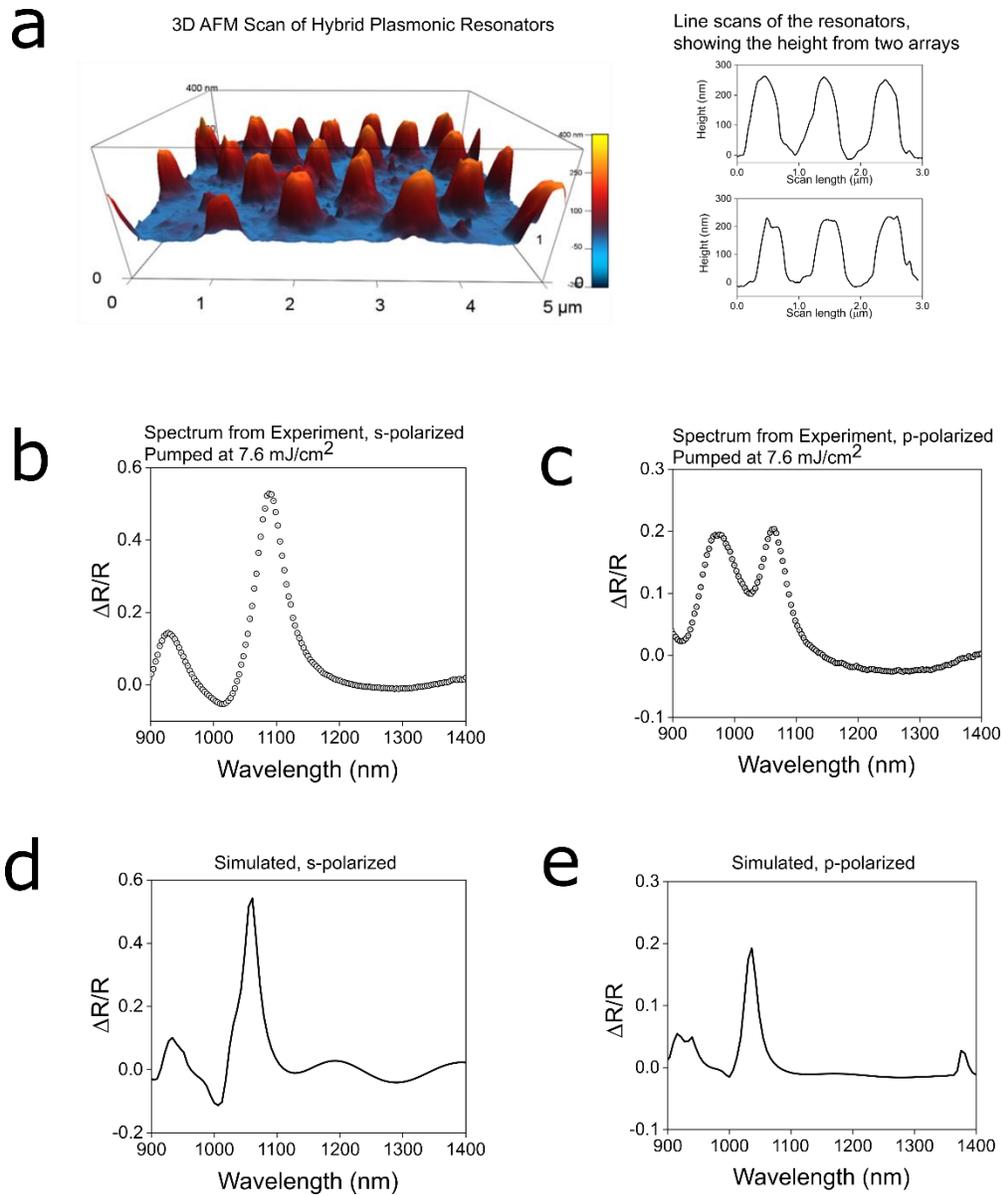

*Figure S7 (a) 3D topology of nanodisks measured by atomic force microscopy (AFM); the height varies from 200 to 250 nm, shown by the line scans. (b) Experimentally measured reflectance modulation for s polarized light and (c) p-polarized light) at a fluence of 7.6 mJ/cm$^2$ (e) Simulated reflectance modulation for s-polarized light and (e) p-polarized light calculated by averaging over the on-off measured 4 randomly selected nanostructures with a height of around 200-220 nm, showing a good qualitative match with experiments*

*Table S7.1. Time constants for the hybrid plasmonic resonator at different pump fluences*

| Pump fluence (mJ/cm$^2$) | Time constant (ps) |
|---|---|
| 0.95 | 2.28 |
| 1.90 | 2.68 |
| 3.80 | 2.00 |
| 7.60 | 1.99 |